\documentclass[letterpaper,english,amsmath,amssymb,reprint]{revtex4-2}
\usepackage[T1]{fontenc}
\usepackage[latin9]{inputenc}
\usepackage{units}
\usepackage{textcomp}
\usepackage{amsmath}
\usepackage{amssymb}
\usepackage{graphicx}

\makeatletter

\pdfpageheight\paperheight
\pdfpagewidth\paperwidth

\providecommand{\tabularnewline}{\\}

\usepackage{url}
\usepackage{hyperref}
\makeatother

\usepackage{babel}
\begin{document}
\preprint{APS/123-QED}
\title{IWAVE - An Adaptive Filter Approach to Phase Lock and the\\
Dynamic Characterisation of Pseudo-Harmonic Waves}
\author{E.\,J.\,Daw}
\email{e.daw@sheffield.ac.uk}

\author{I.\,J.\,Hollows}
\author{E.\,L.\,Jones}
\author{R.\,Kennedy}
\author{T.\,Mistry}
\affiliation{Department of Physics and Astronomy, The University of Sheffield,~\\
 Hicks Building, Hounsfield Road, Sheffield S3 7RH, UK }
\author{T.\,B.\,Edo}
\affiliation{LIGO Laboratory, California Institute of Technology, Pasadena, California
91125, USA, and Department of Physics and Astronomy, The University
of Sheffield,~\\
 Hicks Building, Hounsfield Road, Sheffield S3 7RH, UK}
\author{M.\,Fays}
\affiliation{Department of Astrophysics, Geophysics and Oceanography (GEO), Space sciences, Technologies
and Astophysics Research (STAR), Universit\'e de Li\`ege, all\'ee du six Auot 19, 4000 Li\`ege, Belgium}
\author{L.\,Sun}
\affiliation{OzGrav-ANU, Centre for Gravitational Astrophysics, College of Science,
The Australian National University, ACT 2601, Australia, and LIGO
Laboratory, California Institute of Technology, Pasadena, California
91125, USA\\}
\date{\today}
\begin{abstract}
We present a novel adaptive filtering approach to the dynamic characterisation
of waves of varying frequency and amplitude embedded in arbitrary
noise backgrounds. This method, known as IWAVE\footnote{IWAVE is an acronym for
`Iterative Wave Action-angle Variable Estimator'. The amplitude and phase of the
evolving sinusoid are action and angle variables as described in many textbooks
on classical mechanics.}, possesses critical
advantages over conventional techniques making it a useful new tool
in the dynamic characterisation of a wide range of data containing
embedded oscillating signals. After a review of existing techniques,
we present the IWAVE algorithm, derive its key characteristics, and
provide tests of its performance using simulated and real world data.
\end{abstract}
\maketitle

\section{Introduction}

The co-inventor of the MASER \citep{Schawlow:1958zz}, Arthur Schawlow,
is said to have advised his students: ``Never measure anything but
frequency!'' There exist a great variety of techniques for measuring
the frequencies of pseudo-harmonic waves, broadly addressing two different
classes of problems.

In the first class, the average wave characteristics are estimated
across a measurement interval, under the implicit assumption that
the wave properties are essentially static, or when changes in these
properties over the measurement interval are not of interest. This
problem is addressed by a wide variety of methods, including Welch's
method using discrete Fourier transforms (DFTs) \citep{blackman-tukey,welch_1161901},
Pisarenko's method \citep{Pisarenko:1973}, MUSIC \citep{1143830},
and ESPRIT \citep{1164935}.

In the second class, the oscillator is constantly evolving, and we
seek a time-evolving best estimate of oscillator parameters. This
problem is most often solved using phase locked loops (PLLs) \citep{gardener},
or their hardware realisation, lock-in amplifiers \citep{doi:10.1063/1.1140523}.
The myriad applications of PLLs in science and engineering include, for
example, a recent proposal for digital PLLs as an alternative technology for the readout
of the photodiode clusters used to stabilise the alignment of mirrors in 
future gravitational wave detectors \cite{PhysRevApplied.14.054013}. 

In searches for almost continuous wave (CW) signals in gravitational
wave detectors, the evolving oscillation of continuous wave (CW) signals is at a
very low signal-to-noise ratio (SNR) and so the frequency evolution of such signals
in ground based gravitational wave interferometers cannot be inferred from the raw
data as is necessary for successful tracking with a conventional PLL.
Instead, coherent matched filtering techniques,
such as the $\mathcal{\mathord{F}}$-statistic used in CW searches
\citep{jaranowski1998data,cutler2005generalized}, are able to achieve
higher sensitivity to these weak signals at the price of substantially
greater computational burden. Stack-slide-based semi-coherent algorithms
expedite the computation to some extent, at the cost of sensitivity,
by summing the signal power in multiple coherent segments after sliding
the segments in the frequency domain to account for the signal phase
evolution \citep{brady2000searching,Dergachev05,Mendell05,dhurandhar2008cross}.
More efficient semi-coherent methods, e.g., signal tracking algorithms
based on hidden Markov models \citep{ViterbiA1967Ebfc,1090700,1450960,quinn2001estimation},
have been developed to tackle the computational challenge as well
as to allow for some uncertainties in the signal evolution model \citep{SuvorovaS2016HMmt,sun2018hidden,bayley2019generalized}.

Several other techniques beyond the conventional PLL have been developed
for a range of applications outside the field of gravitational wave research, though
these algorithms are not adapted for the detection of the very weak CW signals
expected from gravitational wave sources. One example is the second order generalised
integrator (SOGI-PLL) \citep{inproceedings}. Developed for the problem of
characterising the characteristics of AC voltages in power lines, SOGI was one of the
first methods to successfully addresses the problem of efficient generation of
a copy of an input sinusoid that is out of phase (the so-called quadrature
or Q phase) with the input signal. Other more recent papers, for example
\citep{rodriguez2010multiresonant,matas2013adaptive,xiao2016frequency,xin2016improved},
have developed the SOGI algorithm for a range of practical applications.
A second example is the enhanced PLL (EPLL) \citep{karimi2012unifying,karimi-ghartemani-epll},
which solves the problem of tracking the amplitude of a harmonic wave
in addition to its frequency. Further PLL developments are well summarised
in \citep{golestan2017single}. 

The IWAVE technique described in this paper is a new type of PLL addressing
the dynamic characterisation of evolving
pseudo-sinusoidal signals. Unlike a conventional PLL, the adaptive
element is a filter rather than an oscillator or counter. IWAVE has
certain advantages over existing PLLs. Firstly, IWAVE produces benign output when
the PLL is unlocked. In the case where the output of the PLL is being 
used to control something, for example, some parameter of a gravitational 
wave detector in a closed loop feedback system, then, in colloquial terms, 
IWAVE does no harm when it isn't working. IWAVE naturally tracks the
amplitude as well as frequency using a single feedback loop, unlike EPLL
which requires two loops internally for the amplitude and the phase.
IWAVE is initialised using a small set of free parameters corresponding
directly to physical oscillator properties - just an initial frequency and a single
time constant; other PLL algorithms typically contain many control parameters that
do not have a clear physical meaning. IWAVE also has the ability to characterise,
simultaneously, multiple oscillations having almost-degenerate frequencies 
using the cross-subtraction method described in Section \ref{subsec:IWAVECROSS}.
This last advantage has led to detailed analysis of almost-frequency-degenerate
violin modes of fused silica suspension wires in advanced LIGO \citep{Cumming_2020}.
Comparisons of IWAVE with the SOGI
and EPLL algorithms are given in Appendix \ref{sec:other_methods}.

As we discuss in Section \ref{subsec:Response-to-noise},
IWAVE as currently implemented is not sensitive
to signals where the ratio of the amplitude of the target wave to the root
mean square (RMS) of
the noise background is less than about 0.3. When applied to broadband
gravitational-wave data, therefore, IWAVE cannot be expected to detect 
gravitational wave CW signals at the strengths anticipated for sources
described in the literature. However, preprocessing steps to divide the
data into narrower frequency bands, thereby significantly reducing the 
RMS of the noise, may yield promising search methods. Studies of these
ideas are under investigation, and will form the subject of future papers.
The potential advantage is that, by using the data itself to track the evolving
frequency of the oscillation, IWAVE may be significantly less reliant on banks
of templates for wave evolution, and hence may require significantly 
less computational resources than existing CW search methods.
For now, however, IWAVE represents a simple, well-characterised and useful technique
for analysing quite complex spaces of evolving oscillators at relatively
low computational burden, which is already being applied to studies of
important background oscillations in gravitational wave data. This, then, is
a methods paper describing how IWAVE works, how it performs, and giving
some examples of that performance on gravitational wave data.

The IWAVE algorithm has many potential applications beyond gravitational
wave science. In the control of brushless electric motors, IWAVE could out-perform
standard vector control methods at low rotation rates where the back-emf in the 
motor windings is weak \citep{s100706901}. In radio communications, IWAVE might be
used to separate closely spaced channels with frequency evolution and
multipath splitting \citep{proakis-salehi}.
In heart magnetometry, IWAVE could be used to estimate Fourier coefficients of
the cardio-magnetic signal using an electrocardiograph signal to modify the phase evolution as
the heartbeat rate evolves \citep{Nakaya_1992}.
In physics IWAVE could be
applied to atomic force microscopy \citep{RevModPhys.75.949}, or as part of an optical squeezing scheme
for the readout of interferometers \citep{PhysRevLett.123.231107}. 
The authors look forward with anticipation 
to seeing what other applications we have not noticed.

This paper consists of: a description of the IWAVE method in Section
\ref{sec:The-IWAVE-Method}; a discussion of the limits of applicability
of IWAVE in Section \ref{sec:limits}; and an overview of certain
other PLL methods in Appendix \ref{sec:other_methods}. Space constraints
have led us to leave out many mathematical steps; a full treatment
of the mathematics can be found at \citep{iwavemathematics}. A software
library implementing IWAVE in C with wrappers into MATLAB and PYTHON/NUMPY
is available on a public git repository here \citep{iwavegit}.

\section{\label{sec:The-IWAVE-Method}The IWAVE Method}

\subsection{The Core Algorithm\label{subsec:The-core-algorithm}}

Before writing down the IWAVE core algorithm, we consider as a starting
point the $\mathcal{{Z}}$-transform \citep{oppenheim-schafer} of
a regularly sampled time series, $x_{p}$,

\begin{equation}
\mathcal{{Z}}_{n}(\Omega)=\sum_{p=-\infty}^{n}x_{p}e^{(p-n)\Omega},\label{eq:1_ztransform}
\end{equation}

where $p$ increases with time, and $\Omega=w-i\Delta$ has real part
$w$ the reciprocal of one $e$-folding for the weighting of previous
samples and imaginary part $\Delta$ equal to the frequency of the
$\mathcal{Z}$-transform component in radians per sample. 

$\mathcal{Z}_{n}(\Omega)$ obeys an iteration equation

\begin{equation}
\mathcal{{Z}}_{n}(\Omega)=e^{-\Omega}\mathcal{{Z}}_{n-1}(\Omega)+x_{n}.\label{eq:2_ztransform_iteration_equation}
\end{equation}

Unit amplitude phasor input, $x_{n}=e^{in\Delta}$, to Equation \ref{eq:2_ztransform_iteration_equation}
results in output $\mathcal{{Z}}_{n}=e^{in\Delta}/(1-e^{-w})$, which
has zero phase shift with respect to the input and a larger amplitude
dependent on $w$. Scaling the input by a factor of $1-e^{-w}$ leads
to an iteration algorithm which passes phasors at frequency $\Delta$
with unit gain and zero phase shift

\begin{equation}
y_{n}=e^{-w}e^{i\Delta}y_{n-1}+(1-e^{-w})x_{n.}\label{eq:e_iwave_core}
\end{equation}

This iteration algorithm is the core of IWAVE. As we shall see, it
responds resonantly at frequency $\Delta$. In the language of signal
processing, Equation \ref{eq:e_iwave_core} is an infinite impulse
response (IIR) filter because it generates its $n^{{\rm th}}$ output
using the current input $x_{n}$, the previous output $y_{n-1}$ and
a two-input, two-output, multi-input, multi-output (MIMO) filter;
$y_{n}$ and $x_{n}$ are, in general, complex variables. We will
also need the real representation of the transfer function for IWAVE
derived from Equation \ref{eq:e_iwave_core} by writing $x_{n}=x_{n}^{R}+ix_{n}^{I}$,
$y_{n}=y_{n}^{R}+iy_{n}^{I},$ and $y_{n-1}=z^{-1}y_{n}$, where $z^{-1}$
is the sample delay operator. In these terms, Equation \ref{eq:e_iwave_core}
can be re-written as
\begin{equation}
\left(\begin{array}{c}
y_{n}^{R}\\
y_{n}^{I}
\end{array}\right)=\begin{bmatrix}H_{11}(z^{-1}) & H_{12}(z^{-1})\\
H_{21}(z^{-1}) & H_{22}(z^{-1})
\end{bmatrix}\left(\begin{array}{c}
x_{n}^{R}\\
x_{n}^{I}
\end{array}\right),
\end{equation}

where the elements of the transfer function matrix are
\begin{equation}
{\bf H}(z^{-1})=\frac{\begin{bmatrix}1-e^{-w}z^{-1}\cos\Delta & -e^{-w}z^{-1}\sin\Delta\\
+e^{-w}z^{-1}\sin\Delta & 1-e^{-w}z^{-1}\cos\Delta
\end{bmatrix}}{\left(\frac{1-2e^{-w}z^{-1}\cos\Delta+e^{-2w}z^{-2}}{1-e^{-w}}\right)}.\label{eq:5_mimo_tf}
\end{equation}

\begin{figure}
\includegraphics[clip,width=1\columnwidth]{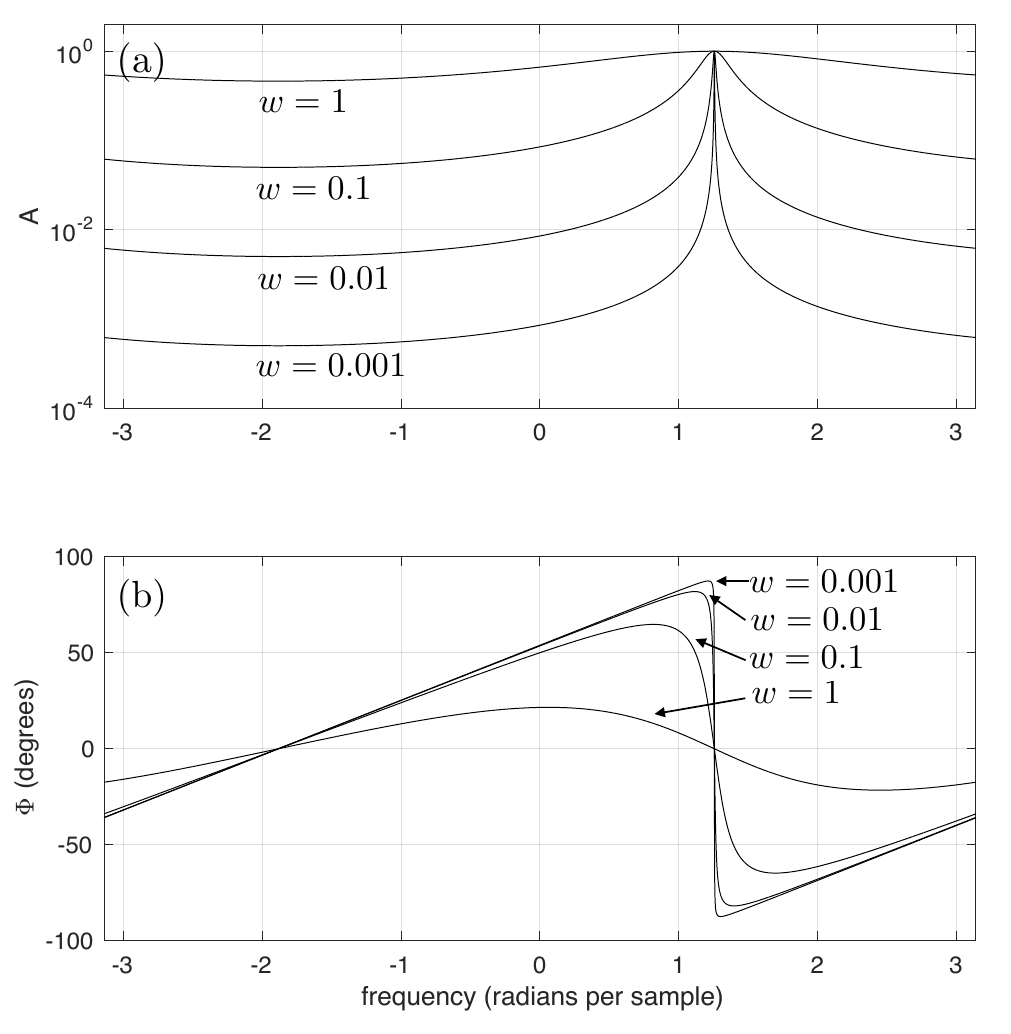}\caption{\label{fig:phasorresp}Magnitude (a) and phase (b) of the response of IWAVE to phasor input as a
function of phasor frequency in radians per sample, for different
values of $w$ and $\Delta=1.257$. A smaller $w$ results in a sharper
resonant peak.}
\end{figure}

In this form, the transfer function is seen to be second order in
sample delay. We determine the response of the core algorithm to an
input consisting of a phasor of arbitrary frequency $\Theta$ radians
per sample by substituting $x_{n}=e^{in\Theta}$ into Equation \ref{eq:e_iwave_core}.
The response is also a phasor, $y_{n}=Ae^{i\left(n\Theta+\Phi\right)}$,
where $A\left(\Theta\right)$ and $\Phi\left(\Theta\right)$ are the
frequency dependent magnitude and phase lag of the output phasor with
respect to the input given by
\begin{equation}
\begin{aligned}A\left(\Theta\right) & =\frac{1-e^{-w}}{\sqrt{1-2e^{-w}\cos\left(\Delta-\Theta\right)+e^{-2w}}}\\
\Phi\left(\Theta\right) & =\arctan\left(\frac{e^{-w}\sin\left(\Delta-\Theta\right)}{1-e^{-w}\cos\left(\Delta-\Theta\right)}\right)
\end{aligned}
.\label{eq:rotfreqresp}
\end{equation}
Thus, phasors of arbitrary frequency are eigenfunctions of the core
algorithm with eigenvalues $Ae^{i\Phi}$. The resonant character of
the eigenvalues at frequency $\Delta$ can be seen in the denominator
of Equation \ref{eq:5_mimo_tf}, which is identical to the resonant
denominator in the SOGI filter \citep{inproceedings}. Figure \ref{fig:phasorresp}
shows A and $\Phi$ as a function of $\Theta$ for ${\rm \Delta}=1.257$
and four values of $w$: $1$, $0.1$, $0.01$, $0.001$. 

Starting from Equation \ref{eq:rotfreqresp} and writing $\delta=(\Delta-\Theta)\ll1,$
and $w\ll1$, so that we are in a limit where the frequency is in
the vicinity of a narrow resonance, the magnitude of the filter output
can be approximated as
\begin{equation}
A(\delta)\simeq\frac{1}{\sqrt{1+\frac{\delta^{2}}{w^{2}}}},\label{eq:lorentzian}
\end{equation}
 so that the peak is approximately Lorentzian in shape with full width
at half maximum (FWHM) of $2w$ radians per sample. Using the sampling
rate, $f_{s}$, in Hz we give other properties of narrow resonances
occurring when $w\ll1$ in Table \ref{tab:iwaveprops}, where $\tau_{s}$
is the sampling period in seconds, $\tau$ is the response time in
seconds and $\Delta_{0}$ is the resonant frequency in radians per
sample.
\begin{table}
\begin{tabular}{|c|c|c|c|}
\hline 
Quantity & Symbol & Formula & Units\tabularnewline
\hline 
\hline 
Full width at half maximum & FWHM or $\Gamma$ & $\frac{wf_{s}}{\pi}=\frac{1}{\pi\tau}$ & Hz\tabularnewline
\hline 
Quality factor & $Q_{f}$ & $\frac{\Delta_{0}}{2w}=\frac{\Delta_{0}\tau}{2\tau_{s}}$ & -\tabularnewline
\hline 
Resonant frequency & $f_{0}$ & $\frac{\Delta_{0}f_{s}}{2\pi}$ & Hz\tabularnewline
\hline 
Response time & $\tau$ & $\begin{array}{c}
\frac{1}{wf_{s}}\\
\frac{\tau_{s}}{w}
\end{array}$ & s\tabularnewline
\hline 
\end{tabular}\caption{\label{tab:iwaveprops}Properties of narrow resonances of the IWAVE
core algorithm in the limit where $w\ll1$. }
\end{table}

\subsection{Application of IWAVE to a Real Sinusoidal Input\label{subsec:Application-of-IWAVE}}

We next discuss the usual case where the input data is a real oscillation
at the IWAVE resonant frequency, $x_{n}=\cos\left(n\Delta\right).$
We decompose $x_{n}$ into two phasors, $x_{n}=x_{f}+x_{b},$ where
$x_{f}=e^{+in\Delta}/2$ and $x_{b}=e^{-in\Delta}/2$, each of which
is an eigenfunction of the core algorithm. Substituting the frequencies
$\pm\Delta$ into Equation \ref{eq:rotfreqresp} and rearranging,
we obtain the response

\begin{equation}
\begin{array}{c}
\begin{aligned}y_{n}=e^{\frac{i\phi}{2}}\left(\frac{(1+a)}{2}\cos\left(n\Delta-\frac{\phi}{2}\right)\right.\end{aligned}
\\
\begin{aligned} & \left.+\frac{i(1-a)}{2}\sin\left(n\Delta-\frac{\phi}{2}\right)\right)\end{aligned}
\end{array}\label{eq:ellipseresp}
\end{equation}
where 
\begin{equation}
\begin{aligned}a & =\frac{1-e^{-w}}{\sqrt{1-2e^{-w}\cos\left(2\Delta\right)+e^{-2w}}}\\
\phi & =\arctan\left(\frac{e^{-w}\sin\left(2\Delta\right)}{1-e^{-w}\cos\left(2\Delta\right)}\right)
\end{aligned}
.\label{eq:backfreqresp}
\end{equation}

\begin{figure}
\includegraphics[width=1\columnwidth]{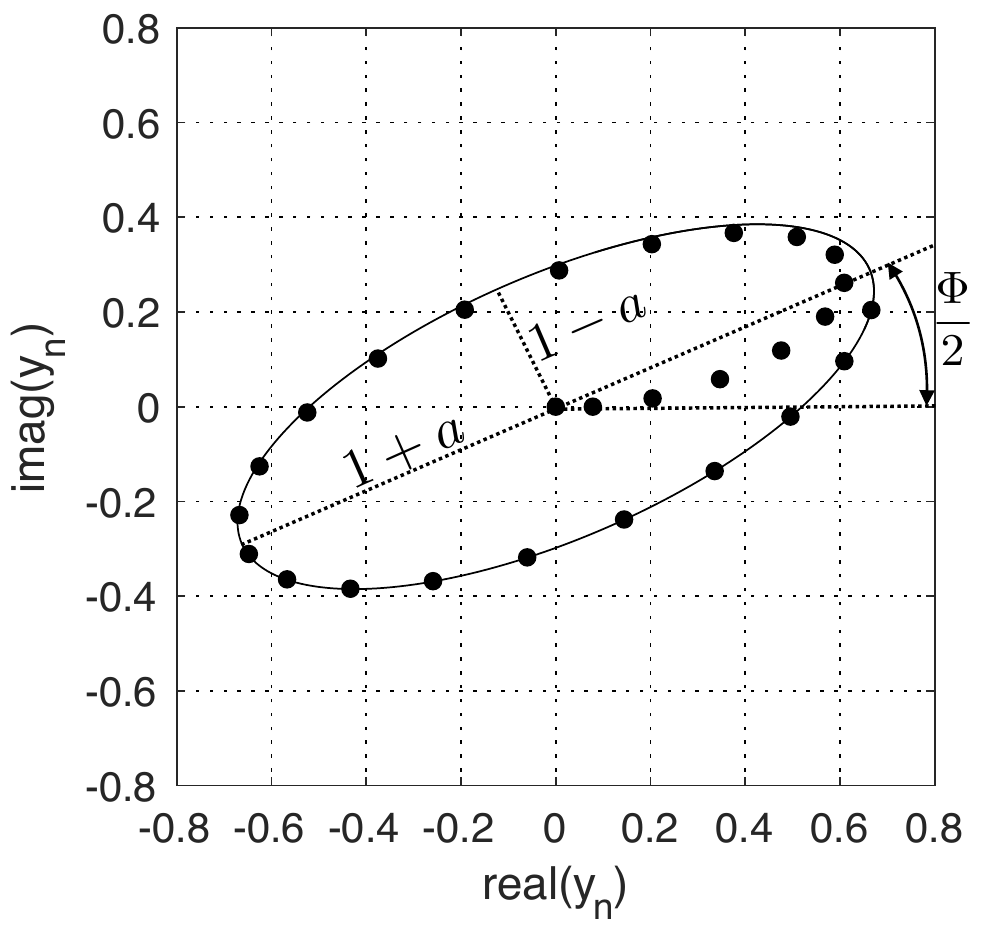}

\caption{\label{fig:phasorout}Output of the IWAVE core algorithm for an input
$x_{n}=\cos n\Delta$ for $\Delta=0.3068,$ $0\protect\leq n<27$
and $w=0.3.$ The points are the individual outputs $y_{n}$ and the
solid line is the limiting ellipse discussed in Section \ref{subsec:Application-of-IWAVE}.}
\end{figure}

By inspection, the locus of $y_{n}$ is an ellipse in the complex
plane having semi-major and semi minor axes $1+a$ and $1-a$. Figure
\ref{fig:phasorout} shows the result of driving the IWAVE core algorithm
with an input $x_{n}=\cos n\Delta$ for $n\geq0$. The output starts
at the origin, spiralling outward towards a limiting ellipse in the
steady state. Notice that the argument of the output $y_{n}$ is always
the same as the phase, $n\Delta$, of the input sinusoid.

Input sinusoids at frequencies other than $\Delta$ also result in
an elliptical steady state output, though with different inclination
angles and eccentricities, and with smaller overall areas due to the
falloff in the magnitude of the response for phasors having frequencies
far from $\Delta$.

A matrix transformation can be used to transform the elliptical locus
into a circular one, with the real part of this circular locus being
a sinusoid in-phase with the input wave, and the imaginary part having
the same amplitude but lagging the input wave by $90^{\circ}$. We
refer to these in-phase and out-of-phase components as the D and Q
phases, respectively. This transformation can be expressed as a sequence
of three elementary operations on the vector whose elements are the
real and imaginary parts of $y_{n}$: a rotation through an angle
of $\frac{\phi}{2}$ about the origin; shears parallel to the real
and imaginary axes by factors of $\frac{2}{1+a}$ and $\frac{2}{1-a}$respectively;
and finally a rotation through an angle of $\frac{-\phi}{2}$ about
the origin. These three matrices can be combined into a single transformation
\begin{equation}
\left(\begin{array}{c}
D_{n}\\
Q_{n}
\end{array}\right)=\left(\begin{array}{cc}
1+e^{-w} & \frac{e^{-w}-1}{\tan\Delta}\\
\frac{e^{-w}-1}{\tan\Delta} & e^{-w}\left(\frac{(e^{w}-1)^{2}}{\sin^{2}\Delta}-1\right)+3
\end{array}\right)\left(\begin{array}{c}
y_{n}^{R}\\
y_{n}^{I}
\end{array}\right).\label{eq:transdq}
\end{equation}

\begin{figure}
\includegraphics[clip,width=1\columnwidth]{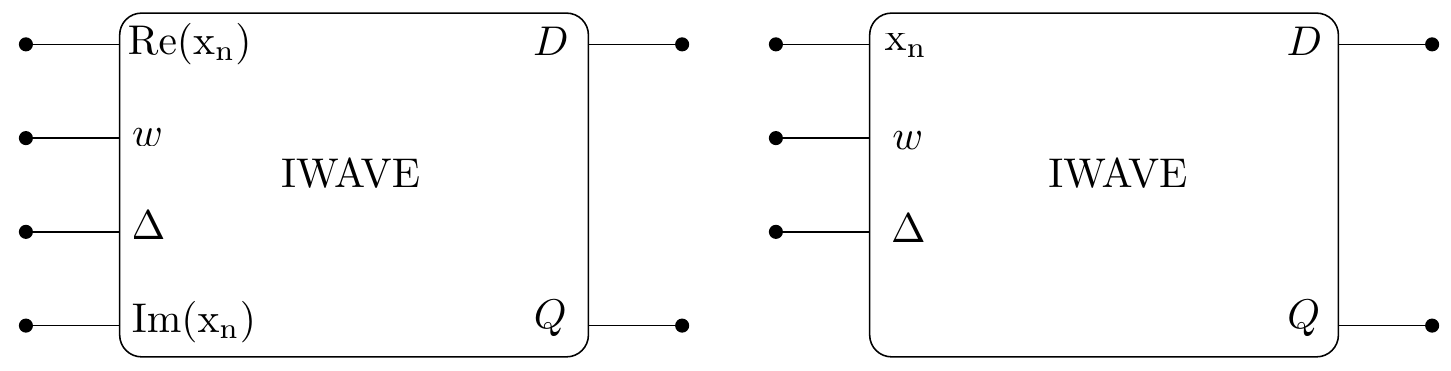}\caption{\label{fig:Symbols-for-IWAVE}Symbols for IWAVE acting on two component
phasor or real inputs. The parameters $w$ and $\Delta$ may be adjusted
to reflect changes in the state of the drive.}
\end{figure}

The IWAVE core algorithm followed by this matrix transformation results
in the output of both D phase and Q phase copies of the input drive.
Thus, IWAVE is an example of what is referred to in signal processing
parlance as an orthogonal state generator. Our convention follows other
papers in the field in that the $D$ ($Q$) phase output is in (out of) phase
with the input at resonance. As we shall see, the Q
phase quadrature can be used to generate an error signal to detect
changes in frequency, allowing IWAVE to be used in place of a reference
oscillator in a PLL. The sum in quadrature of the two phases, $A_{n}=\sqrt{D_{n}^{2}+Q_{n}^{2}}$,
is an estimate of the input signal amplitude. We will use the symbols
in Figure \ref{fig:Symbols-for-IWAVE} to denote the application of
IWAVE either to complex phasor or real sinusoidal inputs. 

\begin{figure}
\includegraphics[clip,width=1\columnwidth]{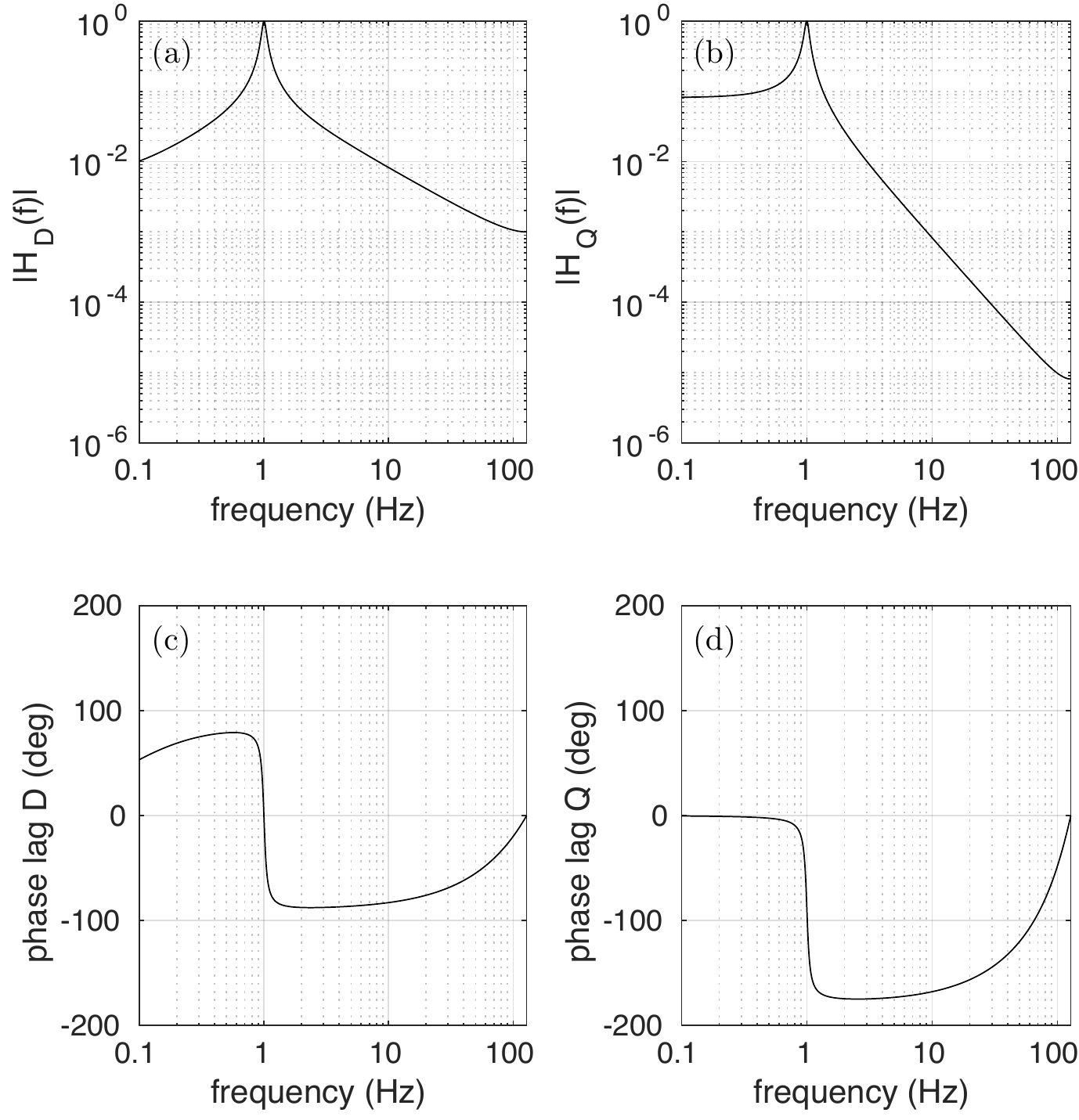}\caption{\label{fig:exampletransferfunctions}An example of the transfer functions
between a real sinusoidal input and the $D$ and $Q$ phase outputs.
Here the sampling rate was 256 Hz, the resonant frequency was 1 Hz
and the filter quality factor $Q$ was set to 12.3. Notice that the
low frequency attenuation in the $Q$ phase output is about 0.08,
which is $1/Q$. Subfigures a and c (b and d) are the magnitude and phase of the D phase
(Q phase) output.}
\end{figure}

We next consider the transfer functions from a real sinusoidal drive,
an example of which is shown in Figure \ref{fig:exampletransferfunctions}.
Note that the dependence on frequency off-resonance is different for
the $D$ and $Q$ outputs. The $D$ transfer function rises linearly
in frequency below the resonance, and falls linearly in frequency
above it, having a phase lead of $90^{\circ}$ below the resonance
and a phase lag of $90^{\circ}$ above it. The $Q$ transfer function
has a flat frequency response below the resonance but falls as $f^{-2}$
above it, and is in phase with the drive below the resonance, but
$180^{\circ}$ out of phase with the drive above it. 

This behaviour is similar to that of a driven series RLC tank circuit,
where the transfer functions from the input voltage to the voltage
across the capacitor and resistor are similar to those between the
input and the $Q$ and $D$ phase outputs, respectively. The relatively
light suppression of low frequency off-resonance signals in the $Q$
phase output can be important, particularly in cases where the quality
factor Q of the circuit is set low by using a relatively large $w$
coefficient. As in the resonant circuit, the ratio of the resonant
to low frequency response is $Q$.

\begin{figure}
\includegraphics[width=1\columnwidth]{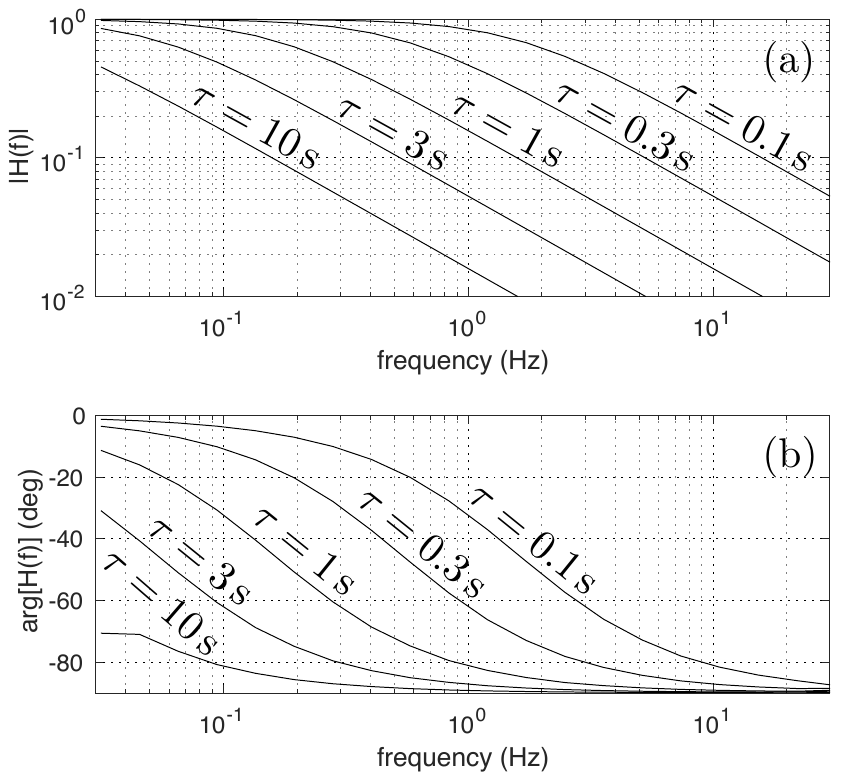}\caption{\label{fig:AMresponse}An example of the response of IWAVE to amplitude
modulated signals. The carrier frequency was $60\,\mathrm{Hz}$ and
the modulation depth was $10\%$. Modulation frequencies in the range
$33\,\mathrm{mHz\protect\leq f_{AM}<33\,\mathrm{Hz}}$ were used and
results are shown for five different values of $\tau$. The 3dB point
for turnover between flat response and proportionality to $1/f_{\mathrm{AM}}$
is $f_{\mathrm{AM}}^{\mathrm{3dB}}=1/\left(2\pi\tau\right)$ in each
case. Subfigures (a) and (b) are the magnitudes and phases of the responses,
respectively.}
\end{figure}

Changes in the amplitude of the incoming wave at the resonance result
in a corresponding change in the quadrature sum, $A_{n}=\sqrt{D_{n}^{2}+Q_{n}^{2}}$,
but with a response time $\tau=\tau_{s}/w$ leading to a single pole
in the response to amplitude changes at $s=-1/\tau.$ Figure \ref{fig:AMresponse}
shows the response at IWAVE's resonance frequency to an input having
a sinusoidally modulated amplitude for a variety of response times.

\subsection{An IWAVE-based Phase Locked Loop}

In order to phase lock with IWAVE, we need a measure of departures
in frequency from the frequency $\Delta$ of the harmonic wave at
the input. We achieve this by exploiting the response time $\tau$
of the IWAVE algorithm. Consider a harmonic wave initially at frequency
$\Delta$. IWAVE yields both a $D_{n}$ phase and a $Q_{n}$ phase
copy of this wave at its output. The product of the out-of-phase copy
and the input wave, $A^{2}\cos\left(n\Delta\right)\sin(n\Delta)=\left(\nicefrac{A^{2}}{2}\right)\sin\left(2n\Delta\right)$,
is a pure harmonic signal at frequency $2\Delta.$ Now, consider an
input signal where the wave develops anomalous phase and amplitude
disturbances, $\delta$ and $\varepsilon$, respectively, so that
$x_{n}=A(1+\varepsilon)\cos(n\Delta+\delta)$. For elapsed times significantly
less than $\tau$ following the onset of these disturbances, the outputs
$D_{n}=A\cos n\Delta$ and $Q_{n}=A\sin n\Delta$ are unaffected by
them. Physically, where the oscillator has an unchanged frequency and 
amplitude, its complex plane representation precesses in a circle about the origin,
and the current IWAVE filter output therefore provides a predictor of the point where
such a static oscillator will land next sample. If the frequency or amplitude evolve,
then the combinations $E_{n}=(x_{n}-D_{n})Q_{n}$ and $F_{n}=x_{n}D_{n}+Q_{n}^{2}-A_{n}^{2}$
are estimates of the departure of the actual evolution of the complex coordinate of the 
oscillator state from this static model. These combinations can be written in matrix form
as follows:

\begin{equation}
\begin{array}{c}
\begin{aligned}\left(\begin{array}{c}
E_{n}\\
F_{n}
\end{array}\right)= & \frac{A^{2}}{2}\left[\left(\begin{array}{c}
-\delta\\
\varepsilon
\end{array}\right)\right.\end{aligned}
\\
\begin{aligned}+ & \left.\left(\begin{array}{cc}
\cos(2n\Delta) & \sin(2n\Delta)\\
-\sin(2n\Delta) & \cos(2n\Delta)
\end{array}\right)\left(\begin{array}{c}
\delta\\
\varepsilon
\end{array}\right)\right].\end{aligned}
\end{array}\label{eq:errorsignals}
\end{equation}

This equation shows that $E_{n}$ and $F_{n}$ each consists of a
static offset plus upper sidebands of frequency $2\Delta$, linear
in the phase and amplitude offsets, respectively. The upper sidebands,
however, appear as components of a rotating phasor in the space spanned
by the $E_{n}$ and $F_{n}$ signals. We remove them using the complex
carrier version of IWAVE, subtracting its outputs from its inputs.
We have determined experimentally that using twice the $w$ factor
in the $2\Delta$ IWAVE filter yields an acceptable error signal for
the detection of phase departures. 

\begin{figure*}
\includegraphics[width=2\columnwidth]{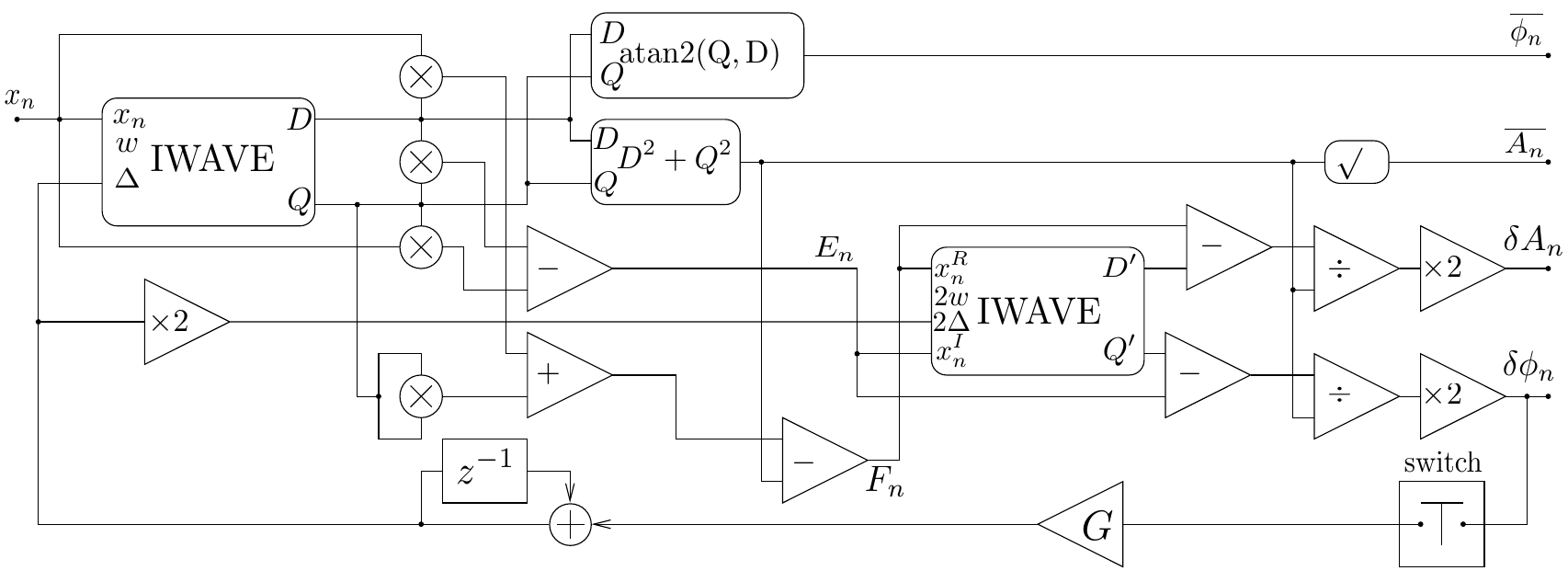}\caption{\label{fig:iwavepllschematic}A schematic of the IWAVE PLL. The error
signal after upper sideband filtering is $\delta\phi_{n}$. The switch
closes the feedback loop to adjust $\Delta$ and $2\Delta$ for the
two IWAVE instances. Use of the complex input IWAVE for attenuation
of the $2\Delta$ component in the error signal reduces the computational
load compared with the use of a real input IWAVE on the $E_{n}$ signal
alone.}
\end{figure*}

Figure \ref{fig:iwavepllschematic} is a schematic for the full IWAVE-based
phase and amplitude detector. We have only considered here the case
where the input data is real and we are tracking harmonic waves. The
case where the input data is two-component rotating phasors is simpler,
as it can be shown that the combination $E_{n}'=x_{n}^{R}Q_{n}-x_{n}^{I}D_{n}$
contains a pure DC signal in phase offset with no upper sideband contamination.
Also shown is the feedback path from the filtered error signal, $\delta\phi_{n}$,
back to $\Delta$ through an integrator, discussed below.

\begin{figure}
\includegraphics[clip,width=1\columnwidth]{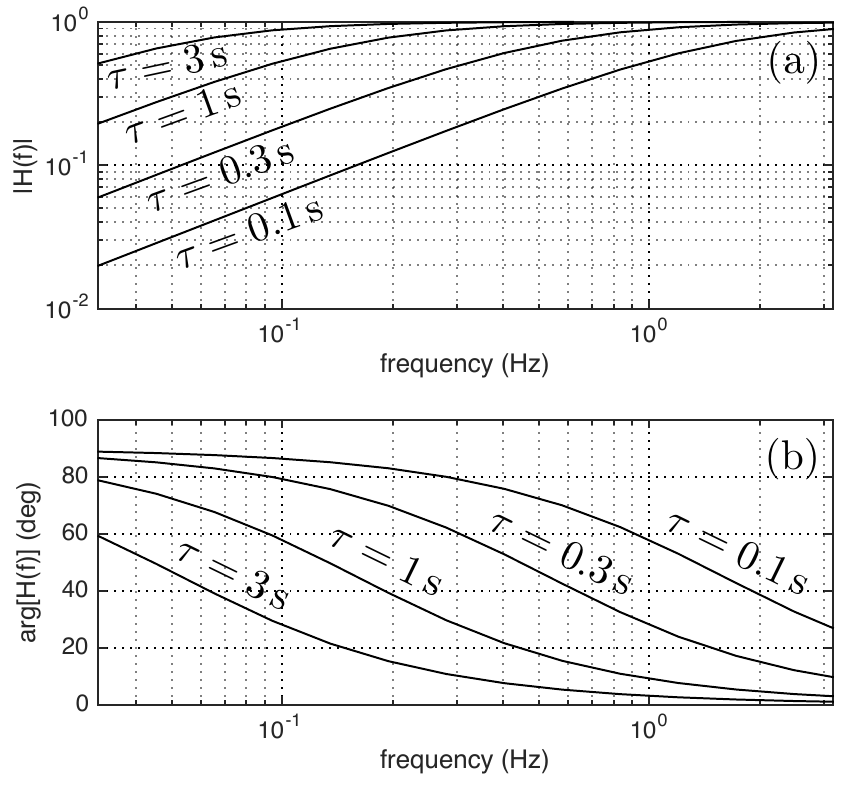}

\caption{\label{fig:fmresp}The transfer function of IWAVE from frequency modulation
of the input carrier to the response in $\delta\phi_{n}$ with the
feedback loop open, for four different values of $\tau$. The output
is insensitive at DC because a step in frequency without feedback
moves the input carrier off the IWAVE resonant frequency $\Delta_{0}$.
Sensitivity increases towards high frequencies because the homodyne
detector relies on beats between the $Q$ phase IWAVE output and the
carrier, which appear so long as IWAVE has not had the adjustment
time, $\tau$, necessary for its outputs to respond to changes in
the input signal. Between these two regimes there is a single pole
at frequency $1/(2\pi\tau)$ Hz, which, combined with the zero at
DC, makes the frequency modulated (FM) response a highpass filter
having the same 3dB point as the lowpass filter associated with the
AM response. In each case the carrier frequency is $\mathrm{60\,Hz}$
and the sampling rate is $\mathrm{16384\,Hz}$. Subfigures (a) and
(b) are the magnitude and phase of the transfer function, respectively.}
\end{figure}

\begin{figure}
\includegraphics[width=1\columnwidth]{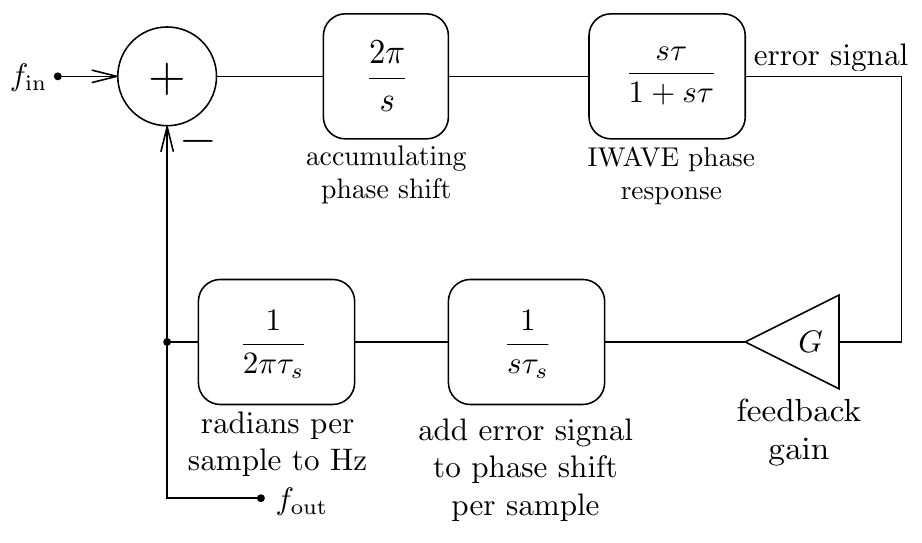}

\caption{\label{fig:iwavepllsplane}An s-plane model of the IWAVE PLL.}
\end{figure}

The response of IWAVE to modulation of the frequency of the carrier
is shown in Figure \ref{fig:fmresp}.
Knowing the analytic form of the frequency response, we can analyse
the closed loop IWAVE PLL to determine an appropriate choice of feedback
gain, $G$. A schematic for the feedback controller is shown in Figure
\ref{fig:iwavepllsplane}.
Any difference between the incoming wave frequency and the IWAVE filter
central frequency results in an accumulating phase shift in the homodyne
detector. This accumulation of phase is represented by the factor
of $2\pi/s$, where $s=2\pi if$, with $f$ being the signal's frequency.
The response of IWAVE to phase, confirmed by the measurements underlying
Figure \ref{fig:fmresp}, acts on the accumulated phase to produce
the error signal. As shown in Figure \ref{fig:iwavepllsplane}, the
feedback path from the error signal to a correction in the central
frequency of IWAVE consists of an adjustable gain, $G$, an addition
of the gain boosted error signal to the previous value of the phase
shift per sample and a scale factor to convert from radians per sample
to frequency in Hz. The closed loop gain is calculated by the usual
consistency argument around the loop,
\begin{equation}
f_{{\rm out}}=f_{{\rm in}}\frac{2\pi}{s}\frac{s\tau}{1+s\tau}G\frac{1}{s\tau_{s}}\frac{1}{2\pi\tau_{s}}-f_{{\rm out}}\frac{2\pi}{s}\frac{s\tau}{1+s\tau}G\frac{1}{s\tau_{s}}\frac{1}{2\pi\tau_{s}},
\end{equation}
from which we obtain the closed loop transfer function,
\begin{equation}
H(s)=\frac{f_{{\rm out}}}{f_{{\rm in}}}=\frac{\frac{G}{\tau_{s}^{2}}}{s^{2}+\frac{s}{\tau}+\frac{G}{\tau_{s}^{2}}}.
\end{equation}
This is the transfer function of a driven damped harmonic oscillator.
We want a critically damped response since the control signal will
be used to measure the oscillator frequency. For critical damping
we require two coincident real poles, which is achieved if $G=\tau_{s}^{2}/(4\tau^{2})$.
The closed loop transfer function then takes the simpler form
\begin{equation}
H(s)=\left(\frac{\frac{1}{2\tau}}{s+\frac{1}{2\tau}}\right)^{2}.\label{eq:closedlooptf}
\end{equation}
The response to frequency or phase modulation is therefore flat below
the knee frequency of $1/(4\pi\tau)\,{\rm Hz},$ where due to the
2 poles it has rolled off to ${\rm -6\,dB},$ and drops as $1/f^{2}$
above that frequency. Larger values of $G$ result in sharper turnover
or a resonant peak, corresponding to the underdamped case.

\section{Limits of applicability of IWAVE\label{sec:limits}}

The time constant, $\tau$, determines the responsiveness of IWAVE to
changes in wave frequency and amplitude, as well as the noise bandwidth
of the filter. Decreasing $\tau$ results in a faster response and
better ability to stay locked on waves whose frequencies are changing,
but also increases the bandwidth of IWAVE to input noise, resulting
a noisier error signal. The optimal $\tau$ is small enough so that
IWAVE stays locked when the frequency changes, but not so small that
excessive background noise is admitted by the filter. Section \ref{subsec:Frequency-tracking}
is on frequency tracking, and Section \ref{subsec:Response-to-noise}
discusses locking in the presence of additive noise and the character
of the error signal.

\subsection{Frequency Tracking\label{subsec:Frequency-tracking}}

\begin{figure}
\includegraphics[scale=0.5]{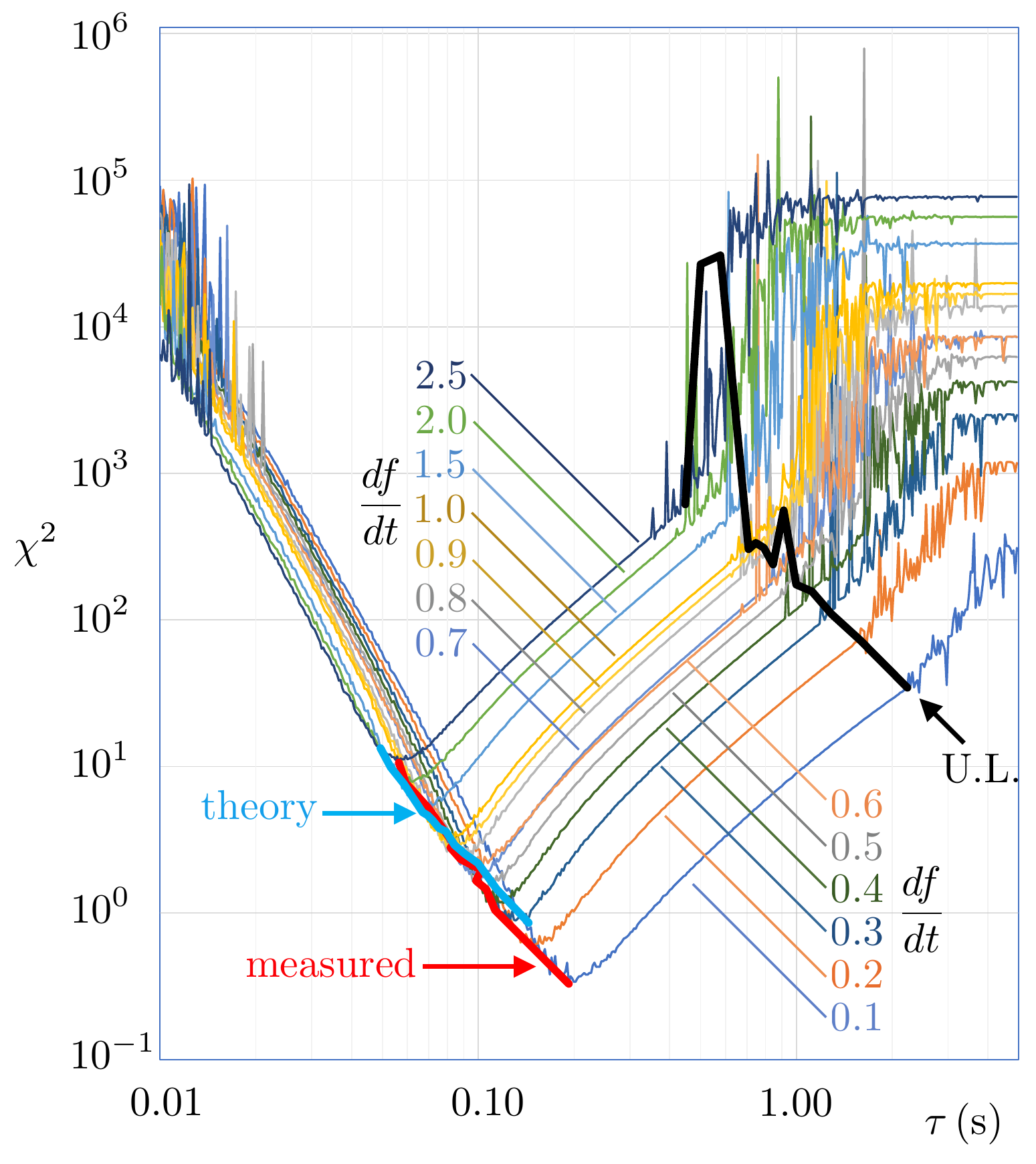}
\caption{\label{fig:ftrack} Measured $\chi^2$ as defined in the 
Section \ref{subsec:Frequency-tracking} from a simulation where
an input swept sinusoid plus additive white gaussian noise was fed
into IWAVE. A range of sweep rates from $0.1$ to $2.5$ $\rm Hz\,s^{-1}$ 
were used. For each injected wave, IWAVE was run with a range of $\tau$
between $\rm 0.01\,s$ and $\rm 5\,s$. Also overlaid are three curves in 
bold. The curve labelled `measured' is the minimum of $\chi^2(\tau)$ 
for each $df/dt$. The curve labelled `theory' is the calculated prediction
for the minimum of $\chi^2(\tau)$ from
Equation \ref{eq:tauopt}. The curve labelled U.L. is the predicted upper limit on $\tau$
above which loss of lock is predicted as discussed in the final paragraph
of Section \ref{subsec:Frequency-tracking}, given by Equation
\ref{eq:taulosspred}, versus
the measured $\chi^2$ at that $\tau$. The thick black line follows the values
of $\tau$ where IWAVE loses lock, as demonstrated by the onset of noise in 
$\chi^2$, over simulations having a variety of $df/dt$ values.}
\end{figure}

Consider a wave whose displacement at time $t$ is \mbox{$h\left(t\right)=A\cos\left(2\pi f\left(t\right)t\right)$},
so that the frequency of the wave is changing. The ability of IWAVE
to track the evolving wave depends on the value of the response time
parameter, $\tau$. Figure \ref{fig:ftrack} shows the results of
running IWAVE on a swept sinusoidal wave starting at 20Hz and increasing
linearly in frequency at a variety of rates. At each sweep rate, a
variety of values of $\tau$ were used. The sum of the squares of
the deviation between the actual frequency and the frequency reconstructed
by IWAVE over a time interval of $20$ seconds, here referred to as $\chi^{2}$,
was calculated as a measure of the accuracy of frequency reconstruction. 

The value of $\chi^{2}$ is seen to be a function of $\tau$, with
a pronounced minimum that is a function of the sweep rate. The $\chi^{2}$
statistic also becomes noisy at both ends of the range of values of
$\tau$. Here we explain the smooth descent and ascent in $\chi^{2}$
on either side of the minimum, and obtain a formula for the optimim
value of $\tau$. We also explain the onset of noise at either end
of the range of $\tau$. 

Starting at the minimum, $\chi^{2}$ rises with $\tau$ because of
the response time of the closed loop transfer function of the servo,
from Equation \ref{eq:closedlooptf}. This transfer function is that
of two RC lowpass filters in series, each having an exponentially
decaying impulse response of timescale $2\tau$. The response time
of the entire transfer function is therefore the time duration of
the autocorrelation of this exponentially decaying function. This
autocorrelation initially rises linearly in time after the impulse,
before reaching a maximum at time $2\tau$ and then decaying exponentially.
The time between the impulse and the point where the exponential decay
reaches $1/e$ of its maximum value is $6\tau$. This causes the frequency
tracking of iwave to lag behind that of the input swept sine wave,
leading to a frequency discrepancy at any given time of $\Delta f_{1}=6\tau(df/dt)$.

Below the minimum of $\chi^{2}(\tau)$, the rise in $\chi^{2}$
with decreasing $\tau$ is explained by the frequency response of
the core IWAVE algorithm, and is best understood by the analogy between
the IWAVE algorithm and the characteristics of a damped harmonic oscillator
discussed in Section \ref{subsec:Application-of-IWAVE}. At higher
values of damping, the frequency response is maximal at a frequency
$f$ below the natural frequency of the undamped oscillator, $f_{0}$
by an amount given by the relationship $f_{0}=\sqrt{f^{2}+1/\left(2\pi^{2}\tau^{2}\right)}$.
This causes a systematic offset between the frequency, $f$, returned
by IWAVE, and the frequency, $f_{0}$, of the input signal. The square
of this frequency difference is an additional contribution to the
$\chi^{2}$ statistic.

By squaring and adding the frequency discrepancies arising from these two effects,
and minimising with respect to $\tau$, making the assumption that $2\pi^2\tau^2(df/dt)^2\gg1$,
we arrive at a value of $\tau$ that minimises the $\chi^2$ statistic for frequency tracking.
\begin{equation}
\tau_{{\rm opt}}=\frac{1}{\sqrt[6]{288\pi^{4}f^{2}\left(\frac{df}{dt}\right)^{2}}}\label{eq:tauopt}
\end{equation}

Figure \ref{fig:ftrack} shows, in bold red the values of $\tau_{{\rm opt}}$ corresponding
to the measured minimum $\tau$ and $\chi^2$ across each of the $\chi^{2}$ curves, vs.
the measured $\chi^2$ at that minimum, along with, in bold blue, the value of $\tau_{\rm opt}$, versus
the value of $\chi^2$ at $\tau_{\rm opt}$. There is good agreement between the 
theoretical $\tau_{opt}$ and the value determined from simulations, within 29\% for 
the largest value of $df/dt$ studied, and within 13\% for the smallest one. At 
larger values of $df/dt$, $\tau_{\rm opt}$ is smaller, so there is 
more broadband noise in the error signal, and a more sophisticated optimisation 
on $\tau$ would lead to a larger optimal value than that predicted by Equation \ref{eq:tauopt}.

We next discuss the breakdown of IWAVE at high values of $\tau$, where
the $\chi^2(\tau)$ curves become noisy, indicating loss
of lock. The following argument leads
to successful prediction of the value of $\tau$ where this breakdown
occurs. Consider a wave whose displacement at time $t$ is $h(t)=A\cos\left(2\pi f(t)t\right)$,
so that the frequency of the wave is changing. Assume that IWAVE is
locked at time $t$, so that the IWAVE output is $m(t)$ and is equal to $h(t)$. At time $t+\tau$,
the wave displacement has evolved to $h(t+\tau)=A\cos\left(2\pi(f+df)(t+\tau)\right)$.
In the time interval $[t,t+\tau]$ the IWAVE output has not had time to respond
to the frequency shift $df$, and therefore takes the form $m(t+\tau)=A\cos\left(2\pi f(t+\tau)\right)$.
The phase shift between the wave and the IWAVE output at time $t+\tau$ is $\Delta\phi\simeq2\pi\tau\,df$.
The error signal for IWAVE is approximated by the integral $\int h(t')m(t')\,dt'$ 
over time interval $[t,t+\tau]$. If the phase shift between the incoming wave
and the IWAVE output over this time interval is greater than $\pi$, the
error signal will undergo a sign change causing loss of lock.
Therefore, a condition for IWAVE to remain
locked is that $\Delta\phi<\pi$, or $\tau\Delta f<1/2$. Writing
$df=\tau\times df(t)/dt$, we arrive at the upper bound that $\tau$ should
obey
\begin{equation}
\tau<1/\sqrt{2df/dt}
\label{eq:taulosspred}
\end{equation}
for IWAVE to remain locked. This line is
drawn on the $\chi^{2}$ curves in Figure \ref{fig:ftrack} in bold
black, labelled U.L., versus the measured value of $\chi^2$ at that value of $\tau$ in 
each simulation. The limit tracks the onset of lock loss, as demonstrated
by large excursions in $\chi^2$, well, across different trial values of $df/dt$. 

\subsection{Response to Noise \label{subsec:Response-to-noise}}

The error signal for the IWAVE PLL is derived from the product of
the input data stream and the Q phase output of the IWAVE filter,
minus the product $DQ$ of the two iwave outputs.
Where the wave frequency is static, 
this subtraction removes the upper sideband component at
frequency $2f$. When the frequency changes, this causes
an additional transient upper sideband component, which is removed
using a second IWAVE filter at frequency $2f$ having a response time
half that of the primary IWAVE filter. Finally, we divide by the square of the
wave amplitude, because both the
amplitude of the incoming wave, and the amplitudes of both IWAVE
outputs scale linearly with the wave amplitude. We need to divide this
scale out otherwise the PLL loop gain will be dependent on the
wave amplitude.

This is a form of homodyne detector. Because the error signal incorporates
the unfiltered wideband input, the error signal incorporates the broadband
noise of the incoming data. The distribution of the error signal therefore
reflects the spectral characteristics of the input data over the full
Nyquist band. If, for example, the incoming data includes
a time domain transient with a broad spectral distribution, this transient
will be reflected in the time history of the error signal from IWAVE.
If the out of band noise is sufficiently large in amplitude, then
IWAVE will lose lock. 

We have determined experimentally that at the optimal choice of response
time, $\tau_{{\rm opt}}$ given in Equation \ref{eq:tauopt}, then
IWAVE will stay locked when the ratio of the wave peak amplitude to
the root mean square noise amplitude exceeds 0.3. Future work to improve
the performance of IWAVE at lower signal-to-noise ratios could involve,
for example, prefiltering the input data to focus on a narrower frequency
band about the frequency of interest for the waves under study. The
noise content of the error signal, which leads to noise also in the
estimate of the IWAVE frequency, is greater at smaller values of $\tau$
where the bandwidth of the IWAVE filter resonance is larger. At sufficiently
small values of $\tau$, the incursion of noise leads again to loss
of lock. This can be seen in Figure \ref{fig:ftrack} for $\tau<0.02$. 
The optimal value of $\tau$ is affected by higher noise levels,
with larger values than that given in Equation \ref{eq:tauopt} becoming
optimal.

The exact value of $\tau$ where lock loss occurs and the effects of
noise on the optimal value of $\tau$ could be determined by
a detailed stochastic differential equation analysis, and is beyond
the scope of this paper. However, a simple scaling argument can be
made. Noise in the error signal leads to a random component being
added to the phase shift per sample. This means that the reconstructed
frequency, which also forms the servo control signal,
contains a component that undergoes a random walk, and hence
grows with the square root of the number of samples. This means that
you might expect the onset of loss of lock to occur at a $\tau$ which
scales as one over the square of the signal to noise ratio, so that
IWAVE works at $\tau$ above a lower limit that goes down by a factor
of two when the signal to noise ratio is enhanced by a factor of $\sqrt{2}$. This
was verified by injecting additive white Gaussian noise on top of
the swept sine wave, and noting that the threshold for lock loss at
low $\tau$ reproduced this predicted behaviour.

In terms of noise in the error signal, the most significant source
is the product of noise in the input data, since this noise enters
the error signal without bandpassing through the IWAVE filter. However,
this broadband noise has an RMS amplitude independent of the amplitude
of the tracked wave. If the tracked wave drops in amplitude, but the
RMS of the broadband noise at the input does not, then the noise component
of the error signal will scale inversely proportional to the amplitude
of the line. This is exactly what is seen when IWAVE is run in practice
on real data. The normalisation of the error signal with the amplitude
squared is necessary to ensure that the feedback loop gain is independent
of the wave amplitude.

The other function of the error signal is to provide an indication
of whether or not the PLL is locked. With a non-stationary RMS, this
is difficult. However, if we scale the error signal by multiplying
it by the amplitude of the wave being tracked, this stabilises the
RMS of the error signal noise at the level of the line amplitude.
If we further divide by the long-term RMS of the input data, we then
obtain a statistic that has a stable RMS of order unity when the servo
is locked. Departures from lock manifest themselves as large transient
spikes of amplitude greater than 10. This effect will be seen in the
discussion of performance on gravitational wave data in Section \ref{sec:Performance-of-IWAVE}.
In particular, in Figure \ref{fig:ligoiwave}, the error signal plotted for each
of the four harmonics in the study has been scaled in this way.

\subsection{Use of multiple IWAVE filters in parallel\label{subsec:IWAVECROSS}}

\begin{figure}
\includegraphics[scale=0.6]{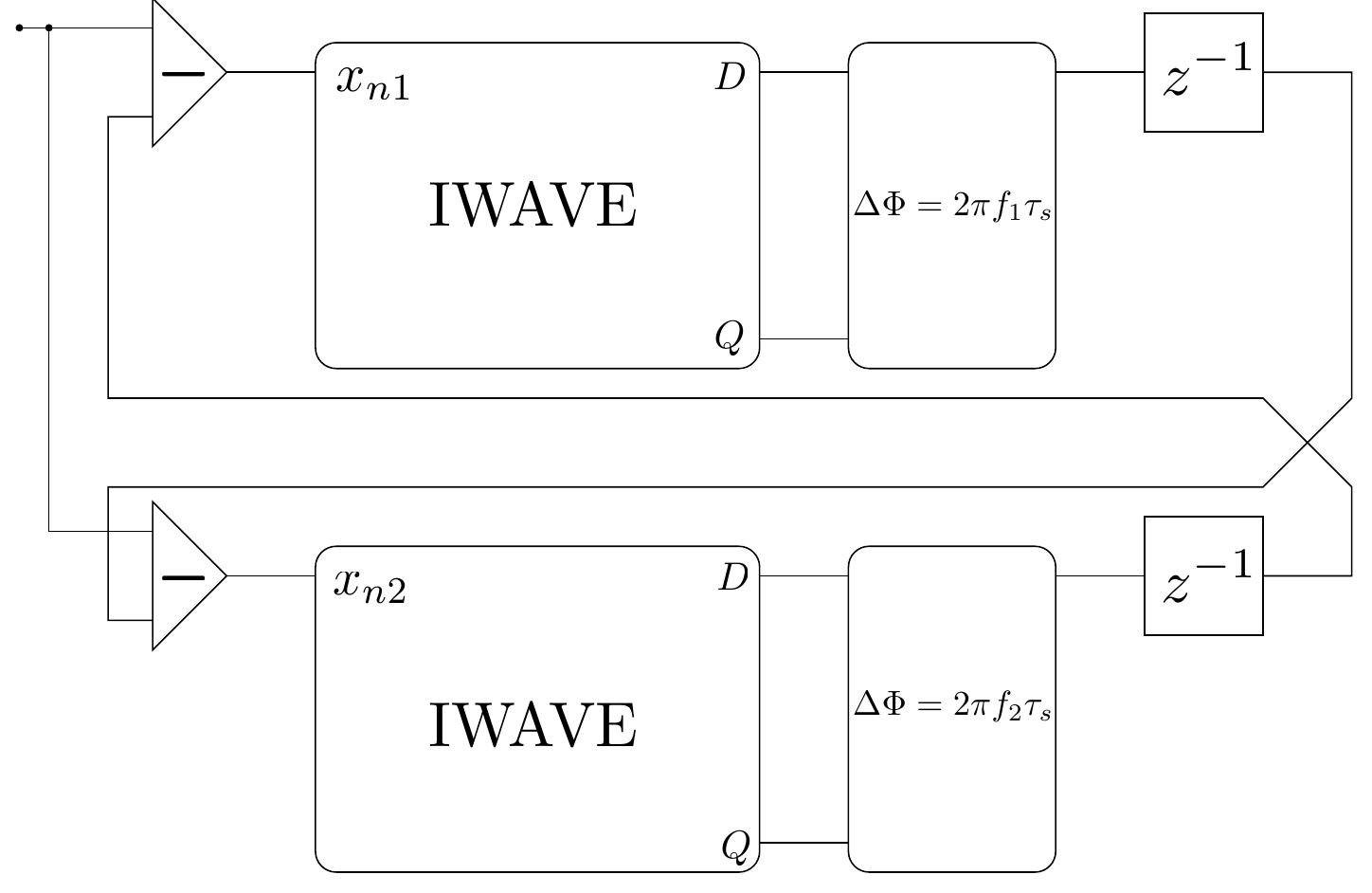}

\caption{A schematic showing the cross subtraction technique for two parallel
IWAVE filters. The feedback loop components are omitted for clarity;
each IWAVE filter is separately instrumented as shown in Figure \ref{fig:iwavepllschematic}.\label{fig:iwavecross}}

\end{figure}

Multiple IWAVE filters can be applied to multiple harmonic waves in
a single data stream. However, when those harmonics are frequencies
spaced closely together, this causes crosstalk between the different
filters. All harmonics present in the data will enter every IWAVE
instance through the error signal. To mitigate this effect we employ
a cross-subtraction scheme, as illustrated in Figure \ref{fig:iwavecross}. 

The schematic shows only two IWAVE instances, but the technique generalises
to any number of filters. Assume that the two IWAVE filters are locked
on line frequencies $f_{1}$ and $f_{2}$. The $D$ and $Q$ phase
outputs from each IWAVE instance are fed into a phase shifter which
estimates the wave signal one sample in the future at the frequency
of the wave tracked by the filter. This wave sample is stored until
the next input data sample, when it is subtracted from the input data
to the other filter. In this way, the input data to each IWAVE filter
is purged of the harmonic being followed by the other filter. This
technique has been determined to successfully lock multiplets of up
to 20 filters, leaving the frequency estimates from each filter free
of oscillations at the difference between frequencies in the multiplet,
the effect seen in the absence of the cross subtraction technique.
The technique works with IWAVE because the two quadrature outputs
can be used to generate an output shifted through an arbitrary phase
shift, and because the outputs are at the same amplitude as the input
wave onto which IWAVE is locked.

\section{Performance of IWAVE on real world data\label{sec:Performance-of-IWAVE}}

We present an example of the application of IWAVE to a set of harmonic
noise components of gravitational wave data taken from the LIGO open
data centre web site \citep{RICHABBOTT2021100658}. The data was acquired
by the LIGO Hanford interferometer on November 30\textsuperscript{th}
2016 and consists of 800 seconds of calibrated strain data from the
Hanford interferometer during an 800 second lock stretch. The data
was preprocessed with a fourth order Butterworth highpass filter at
30Hz, followed by four third order Chebyshev type 2 bandpass filters
between 5Hz and 300Hz, applied in series. Finally, a fine adjustment
to the spectrum was made using a single real pole at 10Hz. The resulting
data is dominated by the 20-80Hz band and is approximately white in
that range. It is not a requirement that the input data to IWAVE be whitened,
but if a spectral feature is to be successfully tracked by IWAVE,
its peak should rise above the noise floor in the surrounding background;
whitening ensures that this is the case. Eight harmonic features were identified from a
broad power spectrum and were tracked using eight parallel IWAVE instances.
These originate from various instrumental sources present in the Hanford instrument
at the time when the data was acquired. Violin mode harmonics have been studied
in detail by Cumming et al. \cite{Cumming_2020}.

\begin{figure}
\includegraphics[scale=0.6]{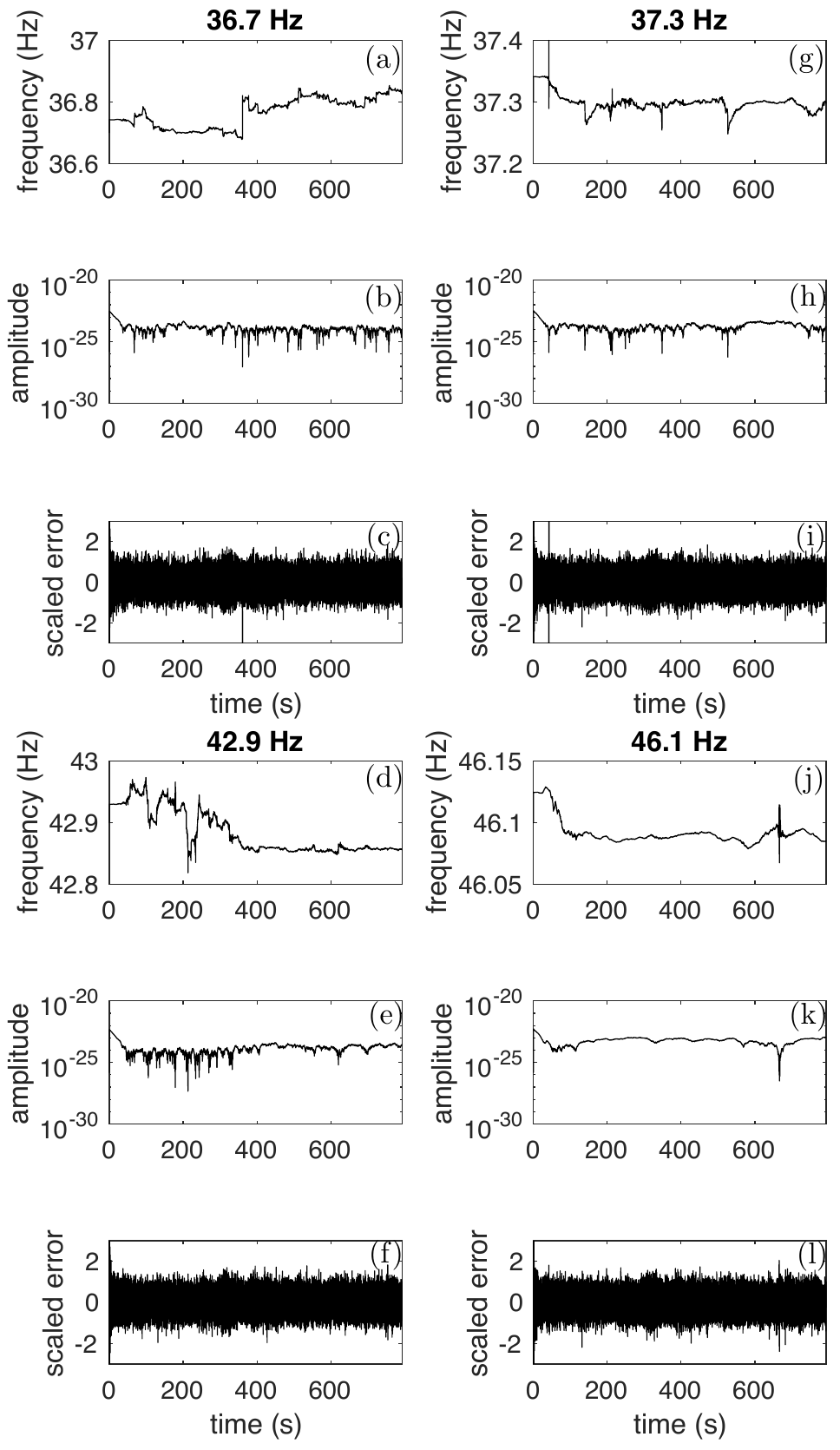}
\caption{The results of IWAVE frequency tracking on four pseudo-harmonics present
in data from the LIGO Hanford detector.  Subfigures a-c, d-f, g-i and h-l pertain
to harmonics at the nominal frequencies displayed at the top of each group of three sub-plots.
At each frequency, the reconstructed frequency, amplitude and scaled error are plotted.
Refer the discussion in Section \ref{subsec:Response-to-noise}
for a description of error signal scaling. Some of the harmonics have frequencies that exhibit
significant time evoultion, others are more static. In a few cases, IWAVE can be seen to have 
lost and re-acquired lock. In the case of the $\rm 36.7\,Hz$ wave, at about $\rm 350\,s$, a lock loss
can be seen in frequency, and also via a spike in the scaled error; similarly at about $\rm 50\,s$ in 
the $\rm 37.3\,Hz$ wave, and at about $\rm 680\,s$ in the $\rm 46.1\,Hz$ wave.}
\label{fig:ligoiwave}
\end{figure}

The results at four of the identified frequencies are shown in Figure
\ref{fig:ligoiwave}. For each harmonic, the frequency, amplitude
and scaled error signal (as described in Section \ref{subsec:Response-to-noise})
are displayed. Amplitudes are in dimensionless strain units, so 1
represents the 4km length of the LIGO detector arms. Though the lines
are in some cases almost degenerate in frequency, there is no evidence
of beats between the reconstructed frequencies. The error signal is
roughly static with approximately unit RMS, though the distribution
is non-Gaussian because of the modulation of the input noise by the
sinusoidal IWAVE output. A non-statistical transient fluctuation in
the modified error signal in the $\rm 36.7\,Hz$ line at around $\rm 360\,s$
corresponds to a jump in the IWAVE reconstructed frequency, indicating
that IWAVE momentarily lost lock when either the frequency of this
sinusoid shifted rapidly or IWAVE jumped between two almost degenerate
harmonics. A second loss of lock can be seen in the $\rm 37.3\,Hz$
data at around $\rm 30\,s$ accompanying a sudden drop in the amplitude
of the harmonic. The eight IWAVE instances all successfully tracked
their target harmonics, with the reconstructed error signal providing
a useful performance indicator.

\section{Summary and future work}

We have described IWAVE, a novel orthogonal state generator, resonant
filter and phase locked loop for the dynamic tracking of harmonic
waves. The method has a single input parameter, its response time.
The algorithm has a low computational load, so that many harmonics
can be tracked in real time using a single CPU core. The ability to
track multiple closely-spaced harmonics means that the method lends
itself well to applications where there are dense `forests' of harmonics,
such as in LIGO violin mode clusters, and communications applications.
IWAVE has been applied to LIGO strain data and used to study the character
of violin modes in ultra low loss fused silica suspensions. There
are many possible applications of the IWAVE method. It is complementary
to existing PLL algorithms that we have described in the Appendix.
We have supplied software implementations of IWAVE in C with MATLAB
and PYTHON wrappers to encourage the community to find other applications
and uses \citep{iwavegit}.

The authors can see several directions in which IWAVE could be improved.
The error signal is susceptible to broadband noise contamination,
and a narrower band alternative would be of benefit in applications
with very weak signals, though narrowbanding will reduce the responsiveness
of the method to frequency changes. There are also applications where
feedback is not important, for example, the use of IWAVE for novel
resonators that is promising for resonant detectors of weak signals
in physics, such as those of dark matter axions \citep{Daw:2018qwb}.
We look forward to seeing what the community finds to do with our
harmonic tracking algorithm.

\section{Acknowledgments}

The members of the Sheffield gravitational wave research group wish
to acknowledge the support of the Science and Technology Facilities
Council under grants ST/V005693/1. ST/V001752/1, ST/V001744/1, ST/V001019/1
and ST/R000336/1. One of us, (I.J.H.) was supported by the Hollows
Scientific Foundation.  L.S. acknowledges the support of the
Australian Research Council Centre of Excellence for Gravitational Wave Discovery (OzGrav), 
Project No. CE170100004, the United States National Science Foundation, and the LIGO Laboratory. 
M.F. acknowledges the support of the Fonds de la Recherche Scientifique-FNRS, Belgium,
under grant No. 4.4501.19.

This research has made use of data, software and/or web tools obtained
from the Gravitational Wave Open Science Center (https://www.gw-openscience.org/
), a service of LIGO Laboratory, the LIGO Scientific Collaboration
and the Virgo Collaboration. LIGO Laboratory and Advanced LIGO are
funded by the United States National Science Foundation (NSF) as well
as the Science and Technology Facilities Council (STFC) of the United
Kingdom, the Max-Planck-Society (MPS), and the State of Niedersachsen/Germany
for support of the construction of Advanced LIGO and construction
and operation of the GEO600 detector. Additional support for Advanced
LIGO was constructed by the California Institute of Technology and
Massachusetts Institute of Technology with funding from the
United States National Science Foundation, and operates under cooperative agreement
PHY-1764464. Advanced LIGO was built under Grant No. PHY-0823459.
Additional support for Advanced
LIGO was provided by the Australian Research Council. Virgo is funded,
through the European Gravitational Observatory (EGO), by the French
Centre National de Recherche Scientifique (CNRS), the Italian Istituto
Nazionale di Fisica Nucleare (INFN) and the Dutch Nikhef, with contributions
by institutions from Belgium, Germany, Greece, Hungary, Ireland, Japan,
Monaco, Poland, Portugal, Spain.

\section{Conflicts of Interest}

The authors have no conflicts of interest to disclose.

\section{Data Availability}

The data that supports the findings of this study are available within this article
and its supplementary material. In the case of the performance data discussed
in Section \ref{sec:Performance-of-IWAVE}, the source data for this analysis can
be obtained from the LIGO open
data centre web site \citep{RICHABBOTT2021100658}.

\appendix

\section{Overview of other PLL methods\label{sec:other_methods}}

\subsection{SOGI-PLL\label{subsec:SOGI}}

The now often-used Generalised Integrator-Based PLL, SOGI-PLL, was
introduced in 2006 \citep{inproceedings}. SOGI is an orthogonal state
generator, producing in-phase and quadrature-phase copies of the input
wave analogous to the IWAVE $D$ and $Q$ outputs. These outputs are
mixed with two quadratures from a reference oscillator, leading to
an error signal in the quadrature phase output that indicates frequency
differences between the SOGI output and the reference. The error signal
is fed to a proportional/integral filter. The filter output is added
to a frequency offset input $\omega_{n}$ and the result is used to
adjust the coefficients of the SOGI filter to reflect the frequency
change, as well as to control the reference oscillator. Essentially,
the SOGI algorithm is coupled to a conventional phase locked loop
with a reference oscillator.

The SOGI algorithm was developed for the field of power grid monitoring,
where frequency changes are a small fraction of a nominal constant
value, commonly ${\rm 50\,Hz}$, ${\rm 60\,Hz}$ or harmonics of these.
It is not designed for large departures from the frequency set at
the $\omega_n$ input. The SOGI orthogonal state generator is designed
using two s-plane integration stages, shown in the tan SOGI block.
These s-plane filters are transformed to the digital domain by, for
example, using a Tustin algorithm. The SOGI method does not track
the wave amplitude, though this can be done in a separate circuit,
assuming the frequency of the wave is known, using homodyne detection,
for example. The SOGI orthogonal state generator does not have the
same transfer functions as the IWAVE one and this is not surprising
given the different methods by which they are obtained, although the
denominators of the two transfer functions are identical, as they
both represent a resonant response to a harmonic drive.

\subsection{EPLL\label{subsec:EPLL}}

The EPLL, although apparently seeming to differ from quadrature signal
generation-based PLLs, such as SOGI-PLL, is closely related thereto
\citep{karimi2012unifying}. A conventional PLL is embedded within
a second feedback servo that uses the integral of the quadrature phase
of the PLL output, multiplied by the unintegrated quadrature phase
to synthesise a $2\omega$ signal, which can then be subtracted from
the input data, compensating for the $2\omega$ component of the phase
detector output inside the PLL. Further, the common amplitudes are
also equal to half the amplitude of the sinusoid in the input data.
In the EPLL structure, the input data are normalised to the PLL by
this amplitude, thereby removing the amplitude dependence of the phase
detector \citep{karimi2013linear}. 

Unlike IWAVE, there is a necessity for two servos to be locked at
once, and several numerical parameters that must be adjusted, including
parameters necessary to implement the s-plane design digitally. EPLL
is susceptible to interference from other waves present in the input
and to DC offsets. The latter two issues are addressed by prefiltering
the input to the EPLL. EPLL has been widely implemented because of
its comparative simplicity, ability to track wave amplitude and the
availability of code implementing the method \citep{gude2019dynamic}.

%\bibliographystyle{apsrev4-2}
%\bibliography{iwave}

\newcommand{\noopsort}[1]{} \newcommand{\printfirst}[2]{#1}
  \newcommand{\singleletter}[1]{#1} \newcommand{\switchargs}[2]{#2#1}

\end{document}